\documentclass[%
 reprint,superscriptaddress,amsmath,amssymb,aps,prl]{revtex4-2}

\usepackage{graphicx}
\usepackage{dcolumn}
\usepackage{bm}
\usepackage{xcolor}
\usepackage[normalem]{ulem}

\begin{document}

\newcommand{\I} 
	{\mathcal I	}
\newcommand{\A} 
	{\mathcal A }
\newcommand{\mP} 
	{\mathcal P	}
\title{Endowing networks with desired symmetries and modular behavior}

\author{P. Khanra} \thanks{These Authors equally contributed to the Manuscript}
\affiliation{Department of Mathematics, University at Buffalo, State University of New York, Buffalo, USA}

\author{S. Ghosh} \thanks{These Authors equally contributed to the Manuscript}
\affiliation{Center for Computational Natural Sciences and Bioinformatics, International Institute of Information Technology, Gachibowli, Hyderabad 500032, India}

\author{D. Aleja}
\affiliation{Universidad Rey Juan Carlos, Calle Tulip\'an s/n, 28933 M\'ostoles, Madrid, Spain}

\author{K. Alfaro-Bittner}
\affiliation{Universidad Rey Juan Carlos, Calle Tulip\'an s/n, 28933 M\'ostoles, Madrid, Spain}

\author{G. Contreras-Aso}
\affiliation{Universidad Rey Juan Carlos, Calle Tulip\'an s/n, 28933 M\'ostoles, Madrid, Spain}

\author{R. Criado}
\affiliation{Universidad Rey Juan Carlos, Calle Tulip\'an s/n, 28933 M\'ostoles, Madrid, Spain}

\author{M. Romance}
\affiliation{Universidad Rey Juan Carlos, Calle Tulip\'an s/n, 28933 M\'ostoles, Madrid, Spain}

\author{S. Boccaletti}
\affiliation{Universidad Rey Juan Carlos, Calle Tulip\'an s/n, 28933 M\'ostoles, Madrid, Spain}
\affiliation{CNR - Institute of Complex Systems, Via Madonna del Piano 10, I-50019 Sesto Fiorentino, Italy}
\affiliation{Moscow Institute of Physics and Technology, Dolgoprudny, Moscow Region, 141701, Russian Federation}
\affiliation{Complex Systems Lab, Department of Physics, Indian Institute of Technology, Indore - Simrol, Indore 453552, India}

\author{P. Pal}
\affiliation{Department of Mathematics, National Institute of Technology, Durgapur 713209, India}

\author{C. Hens} \thanks{Corresponding Author: chittaranjanhens@gmail.com}
\affiliation{Center for Computational Natural Sciences and Bioinformatics, International Institute of Information Technology, Gachibowli, Hyderabad 500032, India}

\date{\today}


\begin{abstract}
Symmetries in a network regulate its organization into functional clustered states.
Given a generic ensemble of nodes and a desirable cluster (or group of clusters), we
exploit the direct connection between the elements of the eigenvector centrality and the graph symmetries to
generate a network equipped with the desired cluster(s), with such a synthetical structure being
furthermore perfectly reflected in the modular organization of the network's functioning.
Our results solve a relevant problem of reverse engineering, and are of generic application
in all cases where a desired parallel functioning needs to be blueprinted.
 \end{abstract}

 \maketitle

Synchronization of networked units is a behavior observed far and wide in natural and man made systems: from brain dynamics and neuronal firing, to epidemics, or power grids, or financial networks~\cite{Breakspear_NatNeuro2017,Strogatz_synchronization_book, Pikovsky_synchronization_book,BoccalettiPhysRep2006, Dorfler_SIAM2012,Motter_NatPhys2013,ashwin2016mathematical,Boccalettibook,Rodrigues_PhysReport2016,Jesus_PRL2007}.
It may either correspond to the setting of a state in which all units follow the same trajectory~\cite{Kundu_EPL2018, Kundu_EPL2018, kundu2020optimizing, Brede_PRE2016}, or to the emergence of structured states where the ensemble splits into different subsets each one evolving in unison. This latter case is known as cluster synchronization (CS)~\cite{dahms2012cluster,Skardal_PRE2011,Nicosia_PRL2013,Ji_PRL2013,sorrentino2007network,williams2013experimental,Pecora_NatCom2014, Sorrentino_SciAdv2016,lodi2020analyzing,bergner2012remote,sorrentino2016approximate,gambuzza2019criterion,Siddique_PRE2018, Sorrentino_SciAdv2016, Cho_PRL2017,Wang_Chaos2019,karakaya2019fading,zhang2017incoherence}, and is the subject of many studies in both single-layer~\cite{Pecora_NatCom2014, Sorrentino_SciAdv2016} and multilayer networks~\cite{DellaRossa_NatCom2020, Sorrentino_IEEE2020}.  Swarms of animals, or synchrony (within sub units) in power grids, or brain dynamics are indeed relevant examples of CS.

The underlying symmetries of a network are responsible for the way nodes split in functional clusters during CS.
In graph theoretic perspective, these clusters are the {\it orbits} of the graph and are the ingredients of the associated symmetry groups.
A {\sl symmetry} (or automorphism) in a graph $G$ is a permutation $\sigma$ of the nodes of $G$ that preserves adjacency, i.e., $\sigma(G)$ is isomorphic to $G$.
If a symmetry $\sigma$ exists such that $\sigma(i)=j$ for some couple of nodes $i,j$, then the two nodes $i$ and $j$ are in the same synchronization's cluster during CS~\cite{Golubitsky1985,Pecora_NatCom2014,Sorrentino_SciAdv2016}.

This leads to the following relevant question: can one design a network of arbitrary number of nodes ($N$) and links ($L$) endowed with an arbitrary set of orbital clusters? In other words, given a desired cluster of nodes (or a group of clusters), can one generate a graph
with density $d = \frac{2L}{N(N-1)}$ endowed with those symmetries which would produce, during CS, exactly the prescribed functional cluster(s)?
A first attempt to solve the problem was offered in Ref. \cite{sorrentino}, where the construction of a feasible quotient graph was proposed as a way to generate networks with prescribed symmetries, a process that implies a noticeable computational complexity and may even be unfeasible for large size networks.

By exploiting the direct connection between the elements of the eigenvector centrality (EVC) and the clusters of a network~\cite{Khanra2021},  we here introduce an effective method able to generate networks with desired and arbitrary sets of nodes, links and clusters, where furthermore the graph structure is perfectly reflected in the modular network's functioning.

\begin{figure}
	\includegraphics[width=0.95\linewidth]{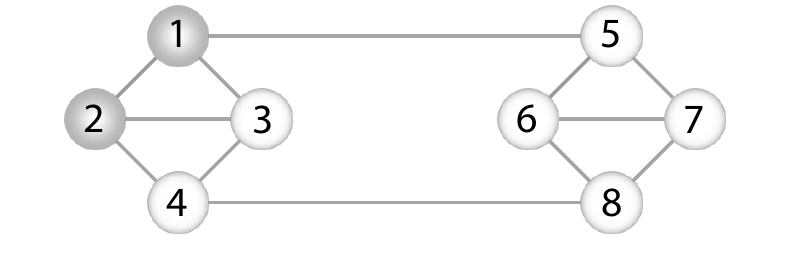}
	\caption{{\bf Symmetries and eigenvector centrality.} Schematic representation of a 8-node 3-regular graph, where all nodes have the same eigenvector centrality but there is no symmetry $\sigma$ such that $\sigma(1)=2$.
	}
	\label{fig:Example}
\end{figure}

We start by recalling that if a symmetry $\sigma$ exists in $G$ permutating nodes $i$ and $j$, then all local invariants (such as the degree, the average distance, etc.) of $i$ and $j$ must be the same. In addition, Ref.~\cite{Khanra2021} demonstrated that $c(i)= c(j)$ (where $c(i)$ and $c(j)$ are the eigenvector centrality of nodes $i$ and $j$).
It should be remarked that the opposite (i.e., $c(i)= c(j)$ implying the existence of a symmetry $\sigma$ such that $\sigma(i)=j$) is not always guaranteed. For instance, the example of Fig.~\ref{fig:Example} is a graph where all the nodes have degree 3, in which $c(i)=c(j)=1/8$ for all pairs $i,j$, and where however there is no symmetry $\sigma$ such that $\sigma(1)=2$, because the average distances from node~$1$ and node~$2$ differ.
At the same time, counterexamples like the one in Fig.~\ref{fig:Example} constitute {\sl pathological cases} limited to regular graph structures, because the construction of symmetries is strongly related with the computation of the isomorphism between graphs~\cite{Mathon1979}. In particular, it has been computationally tested that the EVC is indeed a proper indicator for spotting isomorphic graphs in the case of networks constructed through random processes, for which no realization was found to occur of a graph with two nodes with the same EVC and with no permutating symmetries~\cite{Meghanathan2015}.
Motivated by this evidence, we now move to discuss the methodology for the design of a connected graph $G$ of size $N$ with $k$ prescribed nontrivial clusters.

\begin{figure}
	\includegraphics[width=0.89\linewidth]{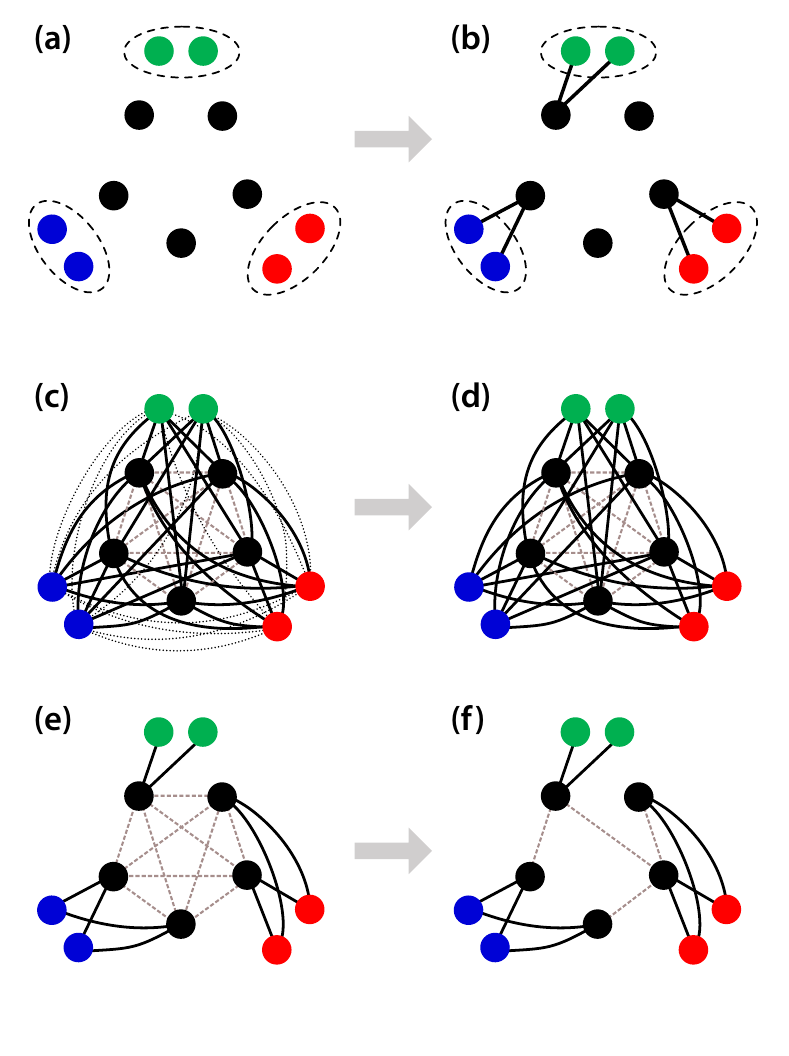}
	\caption{{\bf Network construction.} (a) Three clusters (green, blue and red) are desired. Five black nodes are the trivial (or single node) clusters. (b) Original motifs are formed by connecting nodes within clusters to black nodes: small star graphs are formed where black nodes are the hubs. (c) The original subnetworks are embedded (see text for the procedure). Dotted (dashed) lines denote CC (BB) links (see text for definition) as they bind clustered (unclustered) nodes. Solid lines denote CB links, which have an end in a clustered node and the other in a trivial cluster.  (d) The first step of the specific method consists in removing all CC links. (e) Then, a portion of CB links is judiciously removed to preserve an equal neighborhood for each element of each given cluster. After the removal, different clusters may contain nodes of different degree. (f) Finally, as many BB links as needed are removed to reach the desired network density.}
	\label{Figurereverseengineering}
\end{figure}

For the sake of clarity, we illustrate our method with reference to a small set of $N=11$ nodes, where the goal is to construct a connected network having three nontrivial, desired, clusters of two nodes each (shown with filled green, blue and red circles in Fig.~\ref{Figurereverseengineering} (a), where the black circles represent instead the set of trivial clusters). This is obtained by means of three consecutive steps.

The first step is the creation of sub-networks, or motifs. Here, one considers all nodes of a desired cluster (say, for instance, the green circles) and connects them with a randomly selected trivial cluster (one black circle). A star sub-network is then formed with the black circle as the hub and all the nodes in the cluster as the leaves. Star sub-networks are made in the same way for all other desired clusters [red and blue circles in  Fig.~\ref{Figurereverseengineering}(a)].
The result is the intermediate disconnected network depicted in Fig.~\ref{Figurereverseengineering}(b). By construction, the EVCs of the leaves of each of such sub-networks will be at a same value. If there is not a sufficient number of trivial clusters, one may accomplish this first step by either forming rings or complete graphs with the nodes participating in each individual cluster (in both cases, indeed, the EVC elements corresponding to the nodes of the desired clusters will be the same).
The adjacency matrices for the three star sub-networks [$\A_1$ (green-black), $\A_2$ (blue-black), and $\A_3$ (red-black)] are $
	\A_1=\A_2=\A_3=
	\begin{bmatrix}
	0 & 1 & 1\\
	1 & 0 & 0\\
	1 & 0 & 0
	\end{bmatrix}$.
At the same time, the permutation matrices for each of the star sub-networks are given by
$
		\mP_1=\mP_2=\mP_3=
		\begin{bmatrix}
		1 & 0 & 0\\
		0 & 0 & 1\\
	    0 &1 & 0
		\end{bmatrix}$,
which indeed satisfy $P_i A_i = A_i P_i~(i=1,2,3).$
		
The second step consists in embedding the matrices $\mP_i$s and $A_i$s,
for constructing the permutation ($\mP$) and adjacency ($\A$) matrices of the entire network as
$\mP=
\begin{bmatrix}
		\mP_1 & 0 & 0 & 0\\
		 0 & \mP_2 &  0 & 0\\
		 0 &  0 & \mP_3 &0\\
		 0 & 0 & 0  & \I
\end{bmatrix},   $
and $\A=			
\begin{bmatrix}
				\A_1 & \I & \I & \I\\
				\I & \A_2 & \I & \I\\
				\I & \I & \A_3 & \I\\
				\I & \I & \I & B
\end{bmatrix},
$
where $\I$'s are matrices of appropriate order with unit entries and $B$ is the adjacency matrix corresponding to the trivial clusters. The connected network defined by $\A$ is depicted in Fig.~\ref{Figurereverseengineering}(c), and endowed with the desired clusters.

In summary (and extending the illustration to  generic network of size $N$ with $k$ desired clusters), the first two steps consist in constructing $k$ sub-networks  with corresponding adjacency matrices $\A_i ~(i = 1,\dots, k)$.  For each sub-network ($\A_i$), one then considers the underlying permutation matrix $\mP_i$,  such that $\mP_i \A_i=\A_i \mP_i ~(i = 1,\dots, k)$.  Embedding such units, one ends up with the adjacency matrix
$\A=
	\begin{bmatrix}
	\A_1 & \I & \cdots & \cdots & \I\\
	\I & \A_2 & \I & \cdots & \I\\
	\vdots & \vdots & \ddots & \vdots  & \vdots\\
	\vdots & \cdots & \cdots & \A_k & \I \\
	\I & \I & \cdots &  \I & B
	\end{bmatrix}_{N \times N}$ and the permutation matrix $\mP=
\begin{bmatrix}
\mP_1 & 0 & \cdots &  0  & 0\\
0 & \mP_2 &  &\cdots& 0\\
\vdots &\cdots & \ddots & \cdots  & \vdots\\
\vdots &\cdots & \cdots & \mP_k & 0 \\
0 & 0 & \cdots &  0 & I
\end{bmatrix}_{N \times N}.
$
The resulting network is invariant under the action of $\mP$, but the desired clusters may not be disjoint.

The third step consists in obtaining a network with all desired clusters properly disjoint and with the desired density $d\equiv \frac{2L}{N (N-1)}$. Because of what discussed for the counterexample of Fig.~\ref{fig:Example}, this has to be achieved by a random process of edge removal. Starting from $\A$, the edge removal procedure is as follows: 1) select a given percentage $l$ of edges at random that are not part of the subnetworks $\A_i$; 2) check the connectivity of the network corresponding to the adjacency matrix $A_l$ that results from removing the selected edges from $A$;
3) check the invariance of the resulting network under the action of the permutation $\mP$ i.e.,  $\mP \A_l = \A_l \mP$.
If either step (2) or (3) fails, the edges selected in step (1) are not removed, and a second set of $l$ random links is chosen to test again steps (2-3). Otherwise, the edges are removed, and steps (1-3) are repeated until the desired network density is obtained.
Note that, in this process, one may actually get different networks of the same density with the desired clusters, i.e., the solution of the problem is not unique.

It should be remarked that condition 3) requires checking the permutation invariance ($\mP \A_l = \A_l \mP$) at each step, an operation which may become demanding as the size of the network increases. For networks of arbitrary size, we therefore introduce a more specific
method which takes advantage of the fact that condition 3) is always guaranteed when all the members of each given cluster have the same neighborhood of other network's nodes, so that any two elements of a cluster have the same adjacency.
Looking at Fig.~\ref{Figurereverseengineering}(c), one immediately sees that edges can be divided in three different groups. A first group connect members of different clusters. These links [depicted as dotted lines in Fig.~\ref{Figurereverseengineering}(c)] will be called, from here on, color-color (or CC) links, since they bind nodes of different colors. The second group is made by black-black (or BB) links (dashed lines in the Figure) which have both ends in trivial clusters. Finally, the third group is made by CB (or color-black) links [solid lines in Fig.~\ref{Figurereverseengineering}(c)] which have an end in a element of a cluster and the other end in a trivial cluster.
If $N_k$ is the number of nodes of the $k$-th cluster, $m$ the number of distinct (non trivial) clusters, and $N_m = \sum_{k=1}^{m} N_k$ the total number of clustered nodes,  the initial number of CC, BB and CB links is, respectively, $N_{CC}= \frac{1}{2} \sum_{k=1}^{m} N_k \cdot \left( \sum_{j\neq k} N_j \right)$ , $N_{BB}= \frac{(N-N_m)(N-N_m-1)}{2} $, and $N_{CB}= \sum_{k=1}^{m} N_k \cdot (N-N_m)$.

The first step of the specific method is to remove all the $N_{CC}$ links [see Fig.~\ref{Figurereverseengineering}(d)], which still preserves the feature of same adjacency for each node of each given cluster.
The second step consists in removing judiciously a portion of CB links. While the links forming the original motifs [see Fig.~\ref{Figurereverseengineering}(b)] cannot be removed, if a link is removed connecting a given node of a cluster to a black node, then all the other links connecting all the other nodes of the same cluster to the same black node have to be removed simultaneously, to preserve an equal neighborhood. The result is illustrated in Fig.~\ref{Figurereverseengineering}(e) and one has the additional freedom of imposing a desired degree to each of the clusters (in the example of Fig.~\ref{Figurereverseengineering}(e), green and red nodes end up with having degree 1, while blue nodes have degree 2).
Finally, the third step is removing randomly as many BB links as needed to reach the desired network density (see Fig.~\ref{Figurereverseengineering}(f)). Removing BB links does not affect the neighorhoods of clustered nodes, and therefore permutation is always warranted. The only care here is to check the connectedness of the resulting network. Notice that there is a lower bound for the desired density which is approximately given by  $\bar d = \frac{2 (N_k + (N-N_k-1))}{N(N-1)}= \frac{2}{N}$ (the case where all clustered nodes have degree 1, and the $N-N_k$ trivial clusters form an open ring structure).

Let us now move to show that the cluster organization provided by our method(s) allows the constructed network to behave collectively in the desired modular way.
To this purpose, we first use the exact procedure of the method to design a network of size $N= 1,000$ with two clusters of sizes $20$ and $10$, respectively, and a desired link density $d=0.01$. Then, we investigate CS with such a setup.  We associate each node $i$ to a three dimensional state vector ${\bold x}_i \equiv (x_i,y_i,z_i)$ which obeys the R\"{o}ssler oscillator equations \cite{rossler1976}: ${\dot x}_{i}= -y_i - z_i, \ \ {\dot y}_{i} =  x_i + a y_i + \lambda \sum_{i=1}^{N}  \mathcal{A}_{ij} (y_j-y_i), \ \ {\dot z}_{i}= b + z_i(x_i-c)$,
where dots denote temporal derivatives, the adjacency matrix $\mathcal{A}$ encodes the information of the constructed network, and $\lambda$ is a real parameter
quantifying the coupling strength. The used parameters are $a=0.1$, $b=0.1$ and $c=18$, for which each R\"{o}ssler oscillator develops a chaotic dynamics.
Notice that the coupling term affects only the second variable of each oscillators, a circumstance which determines a class II synchronization scenario
(see the details in Chapter 5 or Ref.~\cite{BoccalettiPhysRep2006}) where complete synchronization is warranted
above a certain threshold ($\bar \lambda$).

\begin{figure}
	\includegraphics[width=1\linewidth]{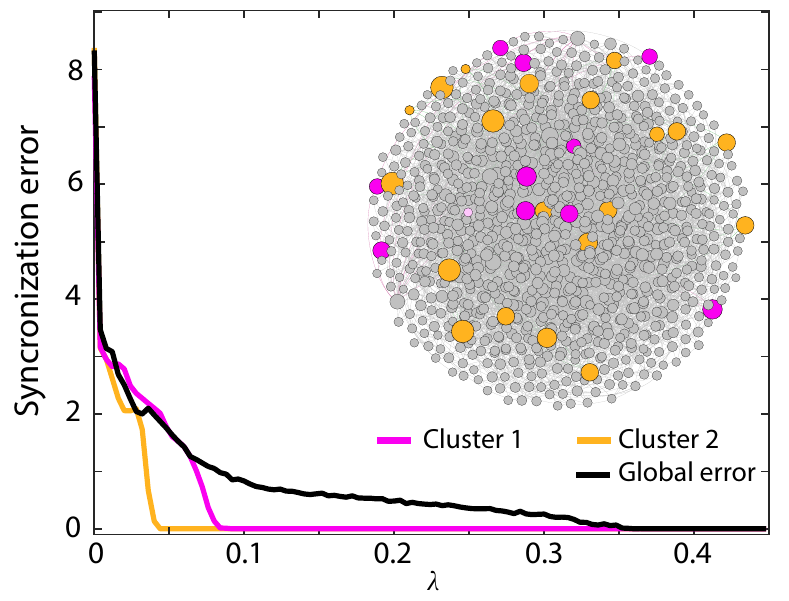}
	\caption{{\bf  Structure induced modular functioning.} Cluster and global synchronization errors (see text for definition and legend for color code) vs. $\lambda$ for the
		constructed network of size $N =1,000$ having two different nontrivial clusters {of $20$ (yellow nodes, degree 4) and $10$ units (magenta nodes, degree 2).} A pictorial sketch of the network is shown in the inset, where gray circles are used to depict all trivial clusters. Notice that nodes in the inset have different sizes only for a better visibility, with no connection with their topological properties. 
	}
	\label{N1000_fig}
\end{figure}

In order to describe what happens for $\lambda < \bar \lambda$, one can monitor the behavior of the $k$-th cluster synchronization errors $S_k$, using the time averaged root mean square deviation defined by
\begin{eqnarray}
S_k=\left \langle \left(\frac{1}{N_k}\sum_{i\in v_k}(y_i-\bar{y})^2\right)^{1/2} \right \rangle_{\Delta T},
\end{eqnarray}
where $v_k$ is the set of nodes contained in cluster $k$,  $\bar{y}$ is the ensemble average of $y$ within the $k$-th cluster, and $\langle . \rangle_{\Delta T}$ denotes temporal average over a time window $\Delta T$ \footnote{Our simulations were performed with a Runge-Kutta fourth order integration algorithm, with integration step $\delta t = 0.01$ time units. Moreover, in each trial, the network was simulated for a total period of 2000 time units, and synchronization errors were averaged over the last $\Delta T = 500$ time units}.
The synchronization errors for the two clusters, as well as the global synchronization error $S_{glob}=\left \langle \left(\frac{1}{N}\sum_{i}(y_i-\bar{y}_{glob})^2\right)^{1/2} \right \rangle_{\Delta T}$ (with $\bar{y}_{glob}$ being the ensemble average of the variable $y$ over the entire network),
are reported in Fig.~\ref{N1000_fig}. The inset in the figure shows a pictorial representation of the constructed network, prepared using the software {\it Gephi},
with the two clusters drawn with different colors (magenta and yellow).
Looking at Fig.~\ref{N1000_fig}, it is seen that $S_{glob}$ decays to zero much later than the synchronization error for the two clusters. Our numerical results show that $S_{glob}$ vanishes at $\lambda=\bar \lambda \sim 0.41$, while the two clusters reach synchronization at different values of $\lambda < \bar \lambda$, so that a large range of coupling strength exists ($0.09 < \lambda < \bar \lambda$) for which the network organizes in a CS state where the two clusters operate in parallel at different synchronized states.

\begin{figure}
	\includegraphics[width=1\linewidth]{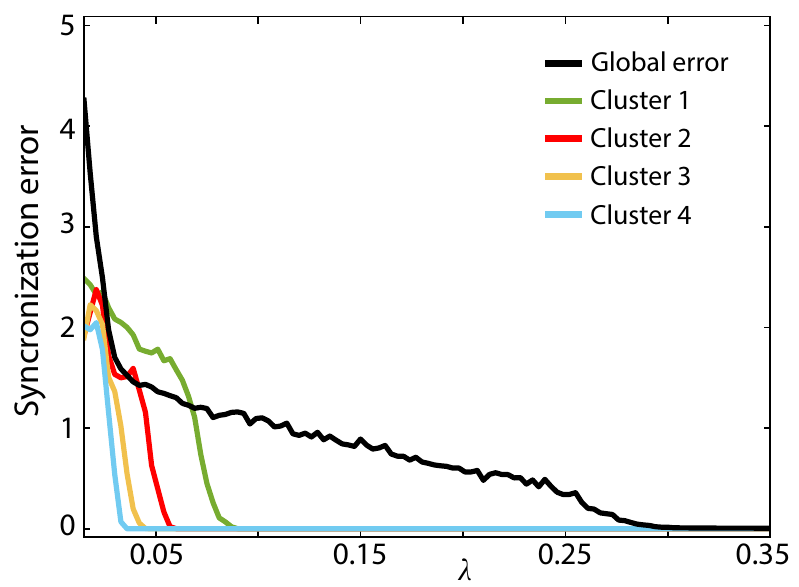}
	\caption{{\bf  Large size networks with differentiated clusters.} Cluster and global synchronization errors (see text for definition) vs. $\lambda$ for a network of size $N =10,000$ where four (very well differentiated in size) clusters have been imprinted by our reduced (specific) method. Namely, clusters 1 to 4 (see color code in the legend) { consist of 1,000 (degree 2), 300 (degree 3), 100 (degree 4), and 30 (degree 5) nodes, respectively. }}
	\label{N10000_fig}
\end{figure}

Finally, we show that our method is effective also when networks have a very large size, as well as when clusters are very well differentiated. For this purpose, we construct a network with $N=10,000$ nodes and density $d= 9.8 \times 10^{-4}$, and we use our reduced (specific) method to blueprint four clusters with sizes spanning more than an order of magnitude.
Namely, clusters 1 to 4 are designed to contain, respectively, 1,000, 300, 100, and 30 nodes. { The degree of a node in each cluster is mentioned in the caption of Fig.\ \ref{N10000_fig}}.
Once again, we associate to each node a vector obeying the R\"{o}ssler oscillator equations \cite{rossler1976} with the same parameters ($a=0.1$, $b=0.1$ and $c=18$) used in Fig. \ref{N1000_fig}. The synchronization errors for the four clusters, as well as $S_{glob}$
are reported in Fig.~\ref{N10000_fig}. Also in this case, one easily sees that the imprinted cluster structure is perfectly reflected by the modular organization of the network's functioning during CS: $S_{glob}$ vanishes at $\lambda=\bar \lambda \sim 0.30$, whereas the four different clusters reach synchronization at different values of the coupling strength in the range $0.03 < \lambda < 0.09$, and therefore a large range of $\lambda$ exists ($0.09 < \lambda < \bar \lambda$) for which the collective network dynamics consists of a CS state with the desired four clusters at works in different synchronized states.
Moreover, the results of Figs. \ref{N1000_fig} and ~\ref{N10000_fig} are beautifully fitting with the analytic predictions given by the Master Stability Approach \cite{Note2master}.

In conclusion, we here solved a relevant problem of reverse engineering, and introduced a method of generic application which allow for the generation of networks with arbitrary number of nodes and links endowed with an arbitrary (and desired) set of orbital clusters, in a way that the graph's parallel functioning occurs into exactly the preselected cluster(s).  This has been accomplished by exploiting the direct connection between the elements of the eigenvector centrality and the clusters of a network.
We then have shown that such a synthetically generated cluster structure is perfectly reflected in the parallel (modular) organization of the network's functioning during cluster synchronization, even for very large sized networks and for clusters well differentiated in size.
As our results are of generic application, and therefore they are of value in a wealth of practical circumstances where networks have to be synthesized and/or generated
with the scope of ensuring a pre-desired parallel functioning.

R. Criado and M. Romance acknowledge funding from projects PGC2018-101625-B-I00 (Spanish Ministry, AEI/FEDER, UE) and M1993 (URJC Grant).  C. Hens is financially supported by the INSPIRE-Faculty grant (Code: IFA17-PH193). G. C-A acknowledges funding from the URJC fellowship PREDOC-21-026-2164.


\begin{thebibliography}{39}
\bibitem [{Note2master()}]{Note2master}%
	\BibitemOpen
	\bibinfo {note} {For class II systems (as it is the present case) Chapter 5 of Ref. 4 establishes that the
 threshold for complete synchronization is nothing but $\frac{\nu*}{\lambda_2}$, where $\nu*$ is the value at which the Master Stability
 Function crosses zero (which, whatever it is, it is the same for the two cases reported in Figs. 3 and 4), and $\lambda_2$
 is the second smallest eigenvalue of the Laplacian matrix. Therefore, this implies a rigorous prediction that the ratio between the two values
 of $\bar \lambda$ at which the global errors vanish in Figs. 3 and 4 should be equal to the reciprocal of the ratio of the two values of $\lambda_2$,
 which is exactly what happens in our simulations}\BibitemShut {NoStop}%
\end{thebibliography}
\providecommand{\noopsort}[1]{}\providecommand{\singleletter}[1]{#1}%

\end{document}